# A New Energy Efficient Routing Algorithm Based on a New Cost Function in Wireless Ad hoc Networks

Mehdi Lotfi, Sam Jabbehdari, and Majid Asadi Shahmirzadi

**Abstract**— Wireless ad hoc networks are power constrained since nodes operate with limited battery energy. Thus, energy consumption is crucial in the design of new ad hoc routing protocols. In order to maximize the lifetime of ad hoc networks, traffic should be sent via a route that can be avoid nodes with low energy. In addition, considering that the nodes of ad hoc networks are mobile, it is possible that a created path is broken because of nodes mobility and establishment of a new path would be done again. This is because of sending additional control packets, accordingly, energy consumption increases. Also, it should avoid nodes which have more buffered packets. Maybe, because of long queue, some of these packets are dropped and transmitted again. This is the reason for wasting of energy. In this paper we propose a new energy efficient algorithm, that uses a new cost function and avoid nodes with characteristics which mentioned above .We show that this algorithm improves the network energy consumption by using this new cost function.

## 1 Introduction

AN ad hoc network is a distributed system consisting of many mobile stations with no predetermined topology and central control. The mobile stations in an ad hoc network communicate wirelessly in a self-organized manner.

Such networks can be used in situations where either there is no other wireless communication infrastructure present or where such infrastructure cannot be used.

Routing in ad hoc networks is one of the most important issues which is discussed in this area, because of unique characteristics of such networks like moving nodes, lack of stable infrastructure, self configuration, and etc. Since all nodes in such networks are mobile means, energy is one of the most important and vital issues for those ones. Routing is one of the issues that discussion of energy has well-deserved influence on it.

There are various parameters in energy efficient routing algorithms that have a role and called cost metrics. It can be referred to some of these metrics such as transmission power of node for sending packets, residual energy of node battery, drain rate of node battery, the number of hops between source and destination, and etc.

Energy efficient algorithms use one or combination of metrics and create cost functions. Then, they use cost functions to select appropriate path between source and destination.

There are numerous and various algorithms, each has own weakness and power.

This paper introduces a new combination of cost metrics and creates a new cost function. Routing algorithm uses this new cost function to select the best path between source and destination. This cost function uses nodes for routing in some way, as a result, sending additional control packets is decreased and then energy consumption of each nodes decreased too.

## 2 Related Works

In conventional routing algorithms like DSDV [1], DSR[2], and AODV [3], which are unaware of energy consumption, connections between two nodes are established through the shortest path route. These algorithms may however result in a quick depletion of the battery energy of the nodes along the most heavily used routes in the network.

In [4], the authors propose an algorithm called Minimum Total Transmission Power Routing (MTPR) using a simple energy metric representing the total energy consumed along a route.

- *Mehdi Lotfi is with the Computer Engineering Department, North Tehran Branch, Islamic Azad University,Tehran,Iran.*
- *Sam Jabbehdari is with the Computer Engineering Department, North Tehran Branch, Islamic Azad University,Tehran,Iran.*
- *Majid Asadi Shahmirzadi is with the Computer Engineering Department, North Tehran Branch, Islamic Azad University,Tehran,Iran.*

Formally, consider a generic route rd $=n_0, n_1, \ldots, n_d$, where $n_0$ is the source node, $n_d$ is the destination node, and $T(n_i,n_j)$ denotes the energy consumed when transmitting over the hop$(n_i,n_j)$, the total transmission power of the route is calculated as:

$$P(r_d) = \sum_{i=0}^{d-1} T(n_i, n_{i+1}) \qquad (1)$$

The optimal route $r_o$ must satisfy the following condition:

$$P(r_o) = \min_{r_j \in r^*} P(r_j) \qquad (2)$$

Where r* is the set of all possible routes. Although MTPR can reduce the total transmission power consumed per packet, it does not reflect the lifetime of each node directly.

Minimum battery cost routing (MBCR) [5] aims to select the route with the minimum aggregate cost/reluctance. Let $c_i(t)$ be the battery capacity of host $n_i$ at time t. One of the possible definitions of f(t), the battery cost function of host $n_i$ is $f_i(t) = 1/c_i(t)$ which is directly related the decrease in residual battery power to the increase in reluctance of the node to participate in routing. The battery cost $R_j$ for route j comprised of $N_j$ nodes is defined as:

$$R_j = \sum_{i=1}^{N_j - 1} f_i(t) \qquad (3)$$

Therefore the best route k satisfies:

$$R_k = \min\{R_j \mid j \in A\} \qquad (4)$$

Where A is the set of all routes under consideration.

Let $f_i(t)$ be a battery cost function for host $n_i$ and $E_i(t)$ the residual battery capacity at a given moment.

The less energy remains in a node, the higher the cost function of this node should be. Authors [6] propose to use $1/E_i(t)$ as cost function. Their Min-Max Battery Cost Routing (MMBCR) metric chooses the path with the least maximal such cost function. In other words, let $r_o$ be the chosen path and r* the set of all possible paths. Then the chosen path fulfills:

$$Cost(R_o) = \min_{r_j \in r_*} \max_{\forall n_i \in r_j} f_i(t) \qquad (5)$$

On one hand, MMBCR considers the weakest node over a path and thereby provides a balanced energy load. On the other hand, there is no guarantee that MMBCR minimizes the total energy consumed over a path.

Conditional Max-Min Battery Capacity Routing (CMMBCR) [6] tries to consider both the minimum transmission energy cost and the maximum network lifetime in the route selection. The CMMBCR presents a hybrid method that selects a route favored by either the MTPR or the MMBCR by using a given threshold γ, which is a percentage value of hosts' initial energy between 0 and 100. When all hosts in possible routes have sufficient remaining battery energy (above the threshold γ×node_s initial energy), a route with the minimum transmission energy cost is chosen. However, if all possible routes have low remaining battery energy (below the threshold), a route with the maximum remaining battery energy is chosen in order to prolong the hosts lifetime.

The FAR protocol [7] assumes a static network and finds the optimal routing path for a given source-destination pair that minimizes the sum of link costs along the path. Here, the link cost for link (i,j) is expressed as:

$$Cost = e_{ij}^{x_1} E_i^{x_2} R_i^{-x_3} \qquad (6)$$

Where $e_{ij}$ is the energy cost for a unit flow transmission over the link and $E_i$ and $R_i$ are the initial and residual energy at the transmitting node i, respectively, and $x_1$, $x_2$, and $x_3$ are nonnegative weighting factors. A link requiring less transmission energy is preferred ($e_{ij}^{x1}$).

At the same time, a transmitting node with high residual energy ($Ri^{-x3}$) that leads to better energy balance is also preferred.

In [8], the authors propose an algorithm that uses the following cost function:

$$Cost(N_i) = (E_c + E_{tx} + E_{rx} + (N-1)E_0) + T.\max[0, E_c + E_{tx} + E_{rx} + (N-1)E_0 - \alpha] \qquad (7)$$

$E_c$ is the used energy at current node, $E_{tx}$ is the energy required for transmitting to the next node, $E_{rx}$ is energy required for receiving from neighbor nodes, $E_o$ is the energy used in overhearing, α is the used energy of a node which has least remaining energy in certain route, and N is the number of neighbors at the current node.

When a node receives RREQ message from others, a receiving node calculates the cost to communicate through that node.

The cost of a route is computed as summary of costs of all nodes that consist of the route and the route that has the least route cost is selected as an optimal route.

## 3 Proposed Algorithm

This section describes used energy model and then our new energy efficient routing algorithm based on a new cost function.

### 3.1 Energy Model

According to IEEE specifications of the network interface card (NIC) with 2 Mbps, the energy consumption varies from 240mA at receiving mode and 280mA in the transmitting mode using 0.5V energy. Thus, when calculating the energy consumed to transmit a packet p is $E(p)=i*v*t_p$ Joules are needed [9] . Here, i is the current, v is the voltage and $t_p$ is the time taken to transmit the packet p. The energy required to transmit a packet p is given by $E_{tx}(p)=280mA*v*t_p$. The energy is required to receive a packet p is given by $E_{rx}(p)=240mA*v*t_p$. The energy consumption of overhearing the data transmission may be assumed as equivalent to energy consumption of receiving of the packet.

### 3.2 New energy efficient routing algorithm based on a new cost function

When a source node has information to send, it sends a Route Request packet. We change Route Request packet in the way that include some additional variables. We use such additional variables to collect necessary information throughout network and make decision about routing. One of these variables is reqSize. Source node puts size of data that wants to send at this variable and sends route request packet. Other additional variables consist of: unstableNodesCount, sumOfNeighbors, and sumOfBufferedPackets, which apply in turn for holding unstable nodes count during path, founding sum of neighbors of all nodes in the path, and founding sum of buffered packets of all nodes in the path. When a node receives Route Request packet from others, a receiving node calculates own remaining lifetime by following equation:

$$RLT_i = \frac{E_i}{DR_i} \qquad (8)$$

$RLT_i$ is the remaining lifetime, $E_i$ is the residual energy, and $DR_i$ is the drain rate of node i and indicates how much the average energy is consumed by a node $n_i$ per second during the interval. The node battery power drain rate actual value is calculated, using the well-known exponential weighted moving average method applied to the drain rate values $DR_{old}$ and $DR_{sample}$ representing the previous and the newly calculated values, as follows:

$$DR_i=\alpha\times DR_{old}+(1-\alpha)\times DR_{sample} \qquad (9)$$

If the node remaining lifetime is more than the needed time to send data packets, which are supposed to send from source to destination, it broadcasts Route Request packet, otherwise it will drop it. In this way, we are preventing participation of nodes, which they finished their energy at exchanging information process and cause to send control packets by other nodes to construct a new path. Node stability or nonstability would be distinguished before broadcasting Route Request. We define stable node as follows:

The node is called stable node that would not change certain rate of its neighbors (50 percent) in specific time (for 2 seconds). If the node is unstable, one adds to unstableNodesCount variable. Also, the number of neighbors and the number of buffered packets add to respective variables amount, and previous amounts of variables would be updated. This manner goes on until Route Request reaches to destination. When Request reaches to the destination, it does not reply immediately. Destination calculates cost of Route Request and buffers it for specific time. If at the same time, it receive another requests with lower cost, they are replaced buffered Route Request, otherwise they will be dropped.

When buffering time expired, buffered Route Request brings out, which has the least cost among received Route Requests and accordingly Route Reply packet is generated and sent. Cost calculation for any Route Request packet is on the basis of cost function as follows:

$$\begin{aligned}&Cost(R) = \\ &w1\times(unstableNodesCount/(hopCount-1)) + \\ &w2\times(sumOfNeighbors/(hopCount-1)) + \\ &w3\times(sumOfBufferedPackets/(hope\ count-1)) \qquad (10)\end{aligned}$$

w1, w2 and w3 are three constant and nonnegative numbers that apply for weighting. Paths are selected with this function, have less unstable nodes, nodes with fewer neighbors, and nodes with fewer buffered packets.

Unstable nodes break the path and cause sending additional control packets, consequently, energy is wasted. Nodes with more neighbors do overhearing more and as a result, lose energy soon. Also, because of multihop routing in ad hoc networks, the probability of passing network traffic via nodes with more neighbors is more than nodes with fewer neighbors. We avoid these nodes and choose nodes with less neighbors, do load balancing and make nodes with less neighbors to participate in network traffic management. Finally, nodes with longer buffer queues cause more timer expiring and consequently packets retransmission because of longer maintenance of packets, and packets retransmission wastes the energy of nodes.

## 4 Simulation Results

We implemented the proposed protocol with glomosim-2.03[10]. Glomosim-2.03 library is scalable simulation environment for wireless network system using the parallel discrete-event simulation capability provided by PARSEC. We tried to compare the performance of our proposed algorithm with MMPR that was implemented by Kwang-Ryoul Kim, Sung-Gi Min and Nam-Kyu Yu. We implemented MMPR on AODV. We used two scenarios for comparison. In first scenario the network size was in the range of [10, 20, 30, 40, 50] nodes and in second one pause time was in the range of [5, 10, 15, 20, 25] seconds. Pause Time in first scenario was fixed at 20s and number of nodes in second scenario was fixed at 30 for the simulation. Other parameters were same in two scenarios and as follows. All network nodes were located in a physical area of size 1000×1000m2 to simulate actual mobile ad hoc networks. The selected mobility model was the Random Waypoint model.

For random waypoint, a node randomly selects a destination from the physical terrain, and then it moves in the direction of the destination in a speed uniformly chosen between the minimum and maximum roaming speed. After it reaches its destination, the node stays there for a specified pause time period. In our simulation, the value of minimum roaming speed was set to 0m/s and maximum mobility speed was 10m/s.

The simulation time of each run lasted for 900 seconds. Each simulation result was obtained from an average of the 10 simulation statistics. The traffic generators were CBR. The generators initiated the first packet (i.e., start time) in different time and sent a 512 bytes packet each time. The values of w1, w2, and w3 were 0.5, 0.3 and 0.2. Buffering time of Route Request by destination node was 100 milliseconds. The initial energy of each node was fixed at 1200J.

As it can be seen in Fig.1 MMPR produced a larger amount of control packets that caused more used energy.

In Fig. 2 the result from proposed algorithm shows the lower used energy compared with MMPR with various numbers of nodes.

Conclusion of Fig. 1 and Fig. 2 is that nodes are increasing while energy consumption in both algorithms increases because of more control packets sending. But energy consumption in MMPR is more than our proposed algorithm because of more control packets sending and more retransmission.

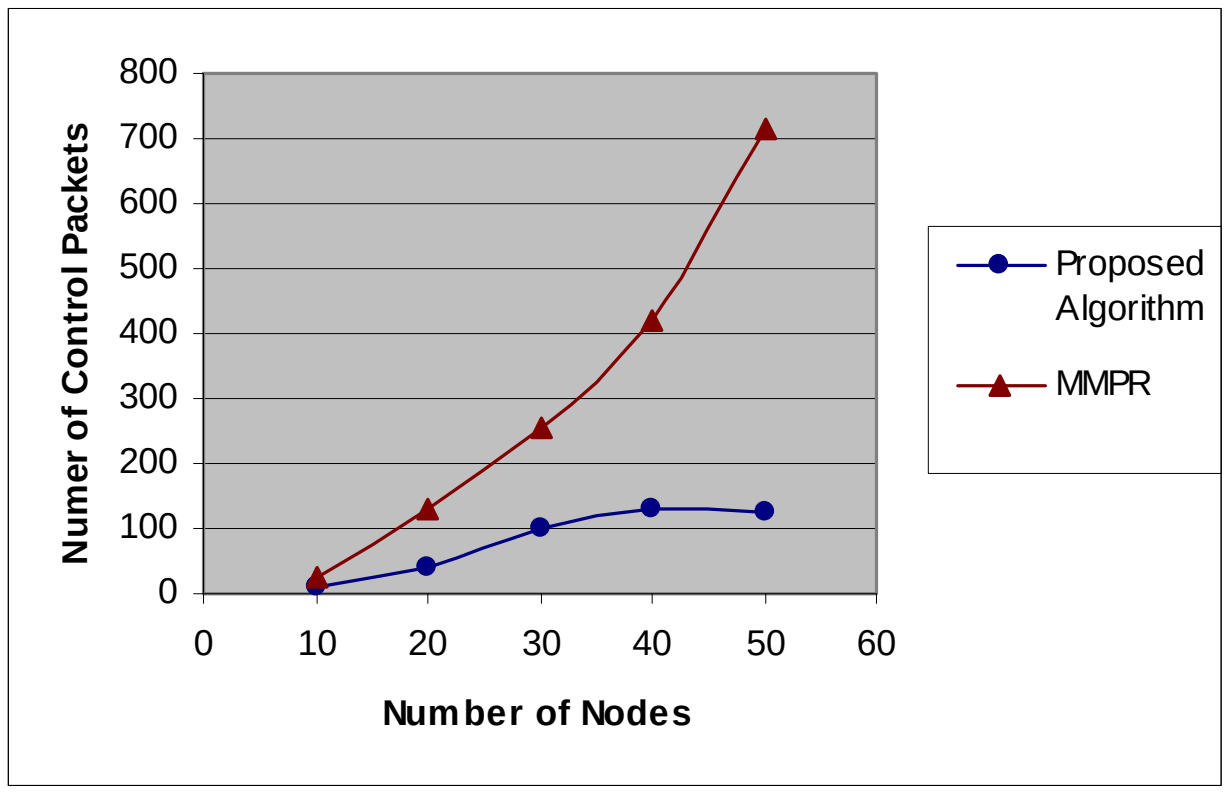

Figure1. Number of Nodes vs. Number of control packets

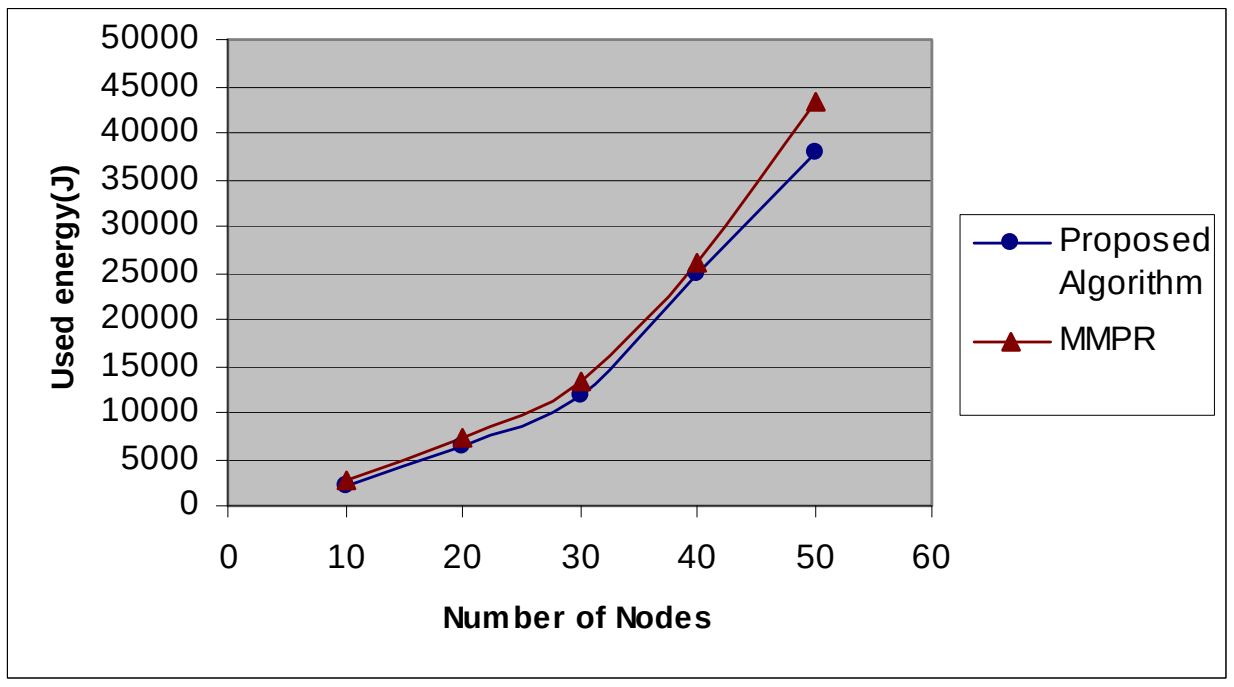

Figure 2.Number of nodes vs. Used energy

Fig. 3 shows the relationship between pause time and number of control packets. Our proposed algorithm sends fewer control packets than MMPR with various pause times.

Fig. 4 shows that our proposed algorithm consumes lower energy in comparison with MMPR. The reason of this issue is fewer control packet sending and fewer retransmission packets.

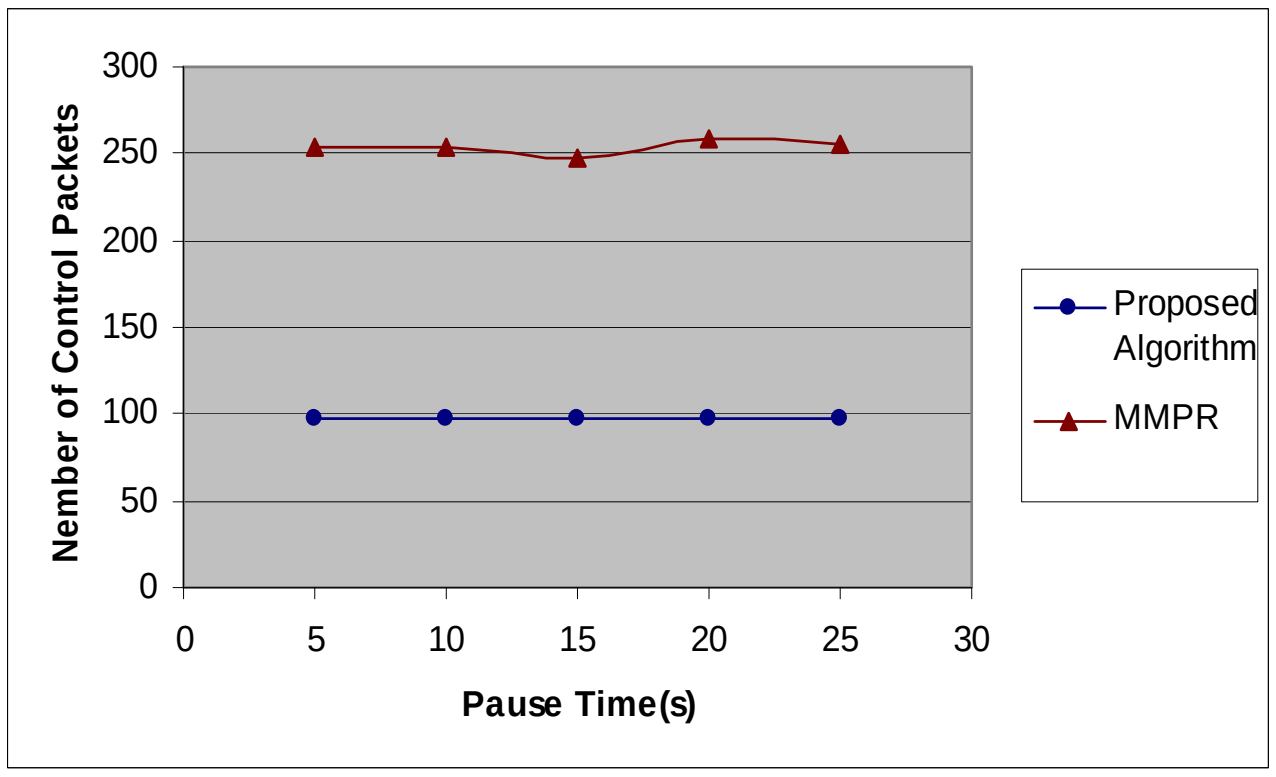

Figure 3.Pause Time vs. Number of Control Packets

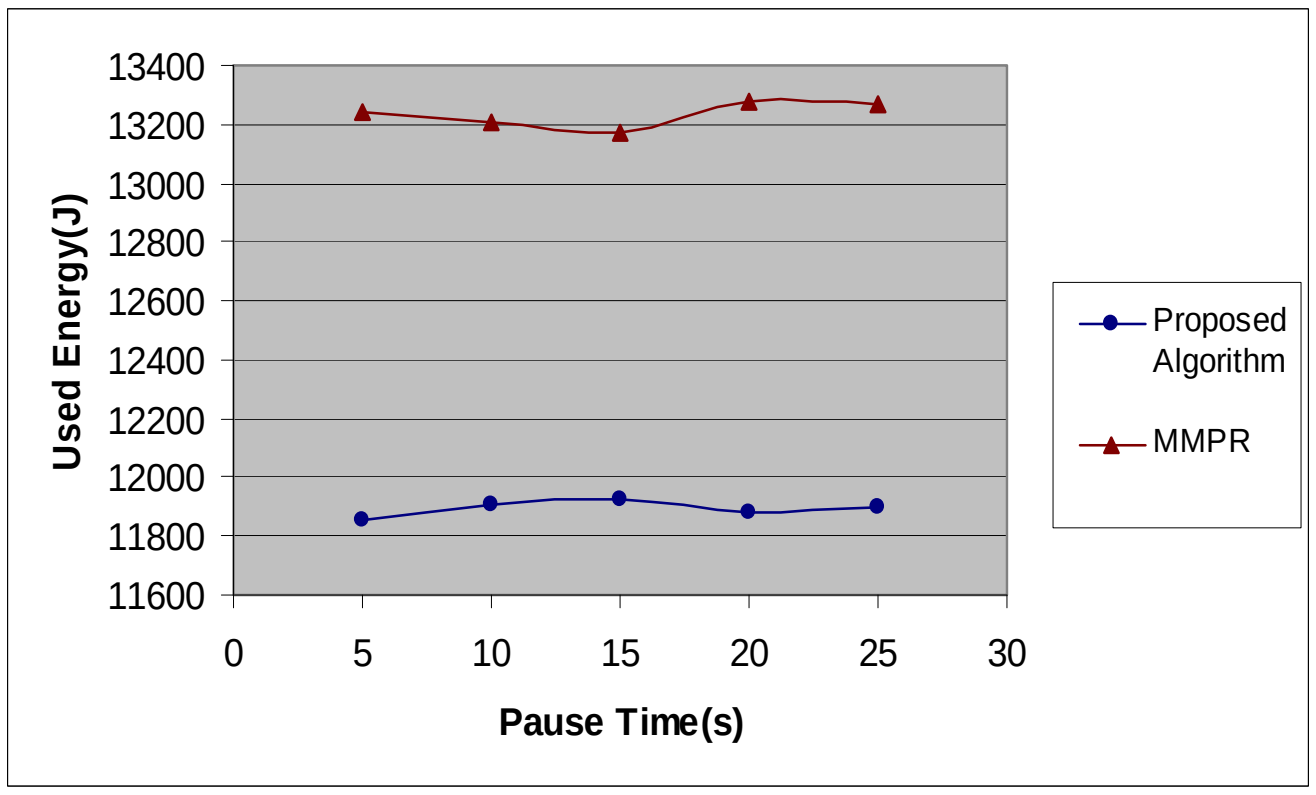

Figure 4.Pause Time vs. Used Energy

## 5 Conclusion

We have discussed a new energy efficient routing algorithm that can be applied to current ad hoc routing protocols such as AODV and DSR. A cost function has been deduced based on nodes stability or nonstability, number of neighbors and buffered packets of nodes and routes are optimized based on the cost functions of nodes. Simulation results have shown that our proposed algorithm improve energy consumption and control packet sending. The energy consumption is balanced among the network and the limited battery resources are utilized efficiently.

In our future work, we want to change our proposed cost function as we use function instead of nonnegative and constant weighting factors: w1, w2, and w3.